\documentclass [preprint]{aastex}      
\usepackage {natbib}
\usepackage[centertags] {amsmath}
\usepackage{hyperref} 

\newcommand{\G}         {\ensuremath{ \textrm{G} }}
\newcommand{\I}         {\texttt{I}\,}
\newcommand{\kpc}       {\textrm{kpc}}
\newcommand{\Msun}      {\textrm{M}\ensuremath{_{\odot}}}
\newcommand{\dex}[1]    {\ensuremath{\times\textrm{10}^{#1}}}

\newcommand{\abs}[1]    {\ensuremath{\mid}{#1}\ensuremath{\mid}}
\newcommand{\sgn}       {\textrm{sgn}}

\renewcommand{\vec}[1]{\mbox{\boldmath$#1$}} 
\renewcommand{\arcsin}   {\ensuremath{\textrm{sin}^{-1}}}

\begin{document}

\title{Potential-density pairs for a family of finite disks}
\author{Earl Schulz}
\affil{60 Mountain Road, North Granby, CT 06060} \email{earlschulz@gmail.com}
\begin{abstract}

Exact analytical solutions are given for the three finite disks with surface density $\Sigma_n=\sigma_0 (1-R^2/\alpha^2)^{n-1/2} \textrm{ with } n=0,
1, 2$. Closed-form solutions in cylindrical co-ordinates are given using only elementary functions for the potential and for the gravitational field
of each of the disks.

The  n=0 disk is the flattened homeoid for which $\Sigma_{hom} = \sigma_0/\sqrt{1-R^2/\alpha^2}$.  Improved results are presented for this disk. The
n=1 disk is the Maclaurin disk for which $\Sigma_{Mac} = \sigma_0 \sqrt{1-R^2/\alpha^2}$.  The Maclaurin disk is a limiting case of the Maclaurin
spheroid. The potential of the Maclaurin disk is found here by integrating the potential of the n=0 disk over $\alpha$, exploiting the linearity of
Poisson's equation. The n=2 disk has the surface density $\Sigma_{D2}=\sigma_0 \left(1-R^2/\alpha^2\right)^{3/2}$.  The potential is found by
integrating the potential of the n=1 disk.

\end{abstract}


\section{Introduction} \label{sec:intro}

Observations of edge-on galaxies now provide detailed structural and kinematic 3-D information. However attempts to use this data to generate mass
distribution have not been completely successful.  One problem is that it is difficult to calculate the potential and force vectors for finite disks.
There is a need for fully solved finite disks which can be used for theoretical studies, for benchmarking computer programs, and as basis functions
to directly model 3-D observations.

Mass modeling commonly assumes that the disk mass in contained in an infinite disk.  These include the exponential disk \citep{fre70}, the Mestel
disk \citep{mes63}, the Kuzmin-Toomre disk \citep{too63,bin08,eva92,con00::1}, and the Rybicki disk \citep{eva93}.

Only a few finite disks have been solved analytically for all $(R,z)$.  \cite{las83} and \cite{vok98} give a solution for a for all $(R,z)$ for a
finite thin disk with constant density.  The gravitational attraction approaches infinity at the edge of this disk. The gravitational attraction of
the finite Mestel disk \citep{mes63,lyn78,hun84} and the truncated exponential disk \citep{cas83} are finite at the edge of the disk but these disks
have not been solved in closed form for points off the disk. \cite{hur08,hur05::1} describe a method to approximate the potential of a power law disk
for points both on and off the disk.

The family of finite disks with surface density $ \Sigma_n(R;\alpha) = \left(1- {R^2}/{\alpha^2}\right)^{n-1/2}$ are related to the Maclaurin
spheroid which has been studied since the time of Newton.  This family has recently been studied by \cite{gon06} and \cite{ped08}.  \cite{gon06} use
the method of \cite{hun63} to obtain the general solution as a sum of Legendre polynomials in elliptical coordinates and give evaluated expressions
for the potentials for the disks 1, 2, and 3. Here we derive complete closed form solutions in cylindrical co-ordinates for the potential and the
gravitational fields of the n=0, 1, and 2 disks. We simplify the integration by moving to the imaginary domain in way which is similar to the
complex-shift method introduced by Appell.  See \cite{cio08,cio07} and references therein

The gravitational potential follows Poisson's equation $\nabla^2\Phi = 4\pi\G\rho$. Poissons's equation is linear so that solutions may be added.
That is, if $\nabla^2\Phi_1=4\pi\G\rho_1$ and  $\nabla^2\Phi_2=4\pi\G\rho_2$  then $\nabla^2(\Phi_1+\Phi_2)=4\pi\G(\rho_1+\rho_2)$. Similarly,
solutions may be differentiated with respect to a parameter to obtain a new solution. \cite{too63} and \cite{lyn89} exploited this linearity to find
a family of velocity-density pairs by differentiating the \cite{kuz56} model .   In the same way,  \cite{sat80} derived new models from the
\cite{miy75} set of potential-density pairs by differentiating with respect to a parameter. Here we use a similar approach by integrating a known
solution with respect to a parameter. If the integral is tractable the result is a new potential-density pair.

Maple 11 was used in this work. Maple was invaluable in simplifying the unwieldy intermediate results but needed a good deal of coaxing to handle the
messier equations. The Maple results were cross checked in several ways.

A common notation is used throughout.   Cylindrical co-ordinates $(R,z)$ are used; $\Sigma$ is the disk surface density; $\sigma_0$  is the surface
density at $R=0$; and $\alpha$ is the disk radius. $\Phi$ is the potential where $\Phi$ is always negative and $\Phi(\infty) =0$;  $F_R$ and $F_z$
are the gravitational field vectors with the standard sign convention, ie: $\vec{F} = - \vec{\nabla} \Phi  $~.

The principal values of elementary functions are used so that, for instance, for $z=x+\I y=r \cos(\theta) + \I  r  \sin(\theta)$;
~~$\sqrt{z}=\sqrt{r}\cos(\theta/2) +\I \sqrt{r}\sin(\theta/2)$, valid for $-\pi<\theta<\pi$.  In this way the function $\sqrt{z}$ is unambiguous with
continuous derivatives except on the negative real axis where it is discontinuous.

\section{The n=0 Disk: The Flattened Homeoidal Shell}

A homeoid is a shell of uniform density which is bounded by similar spheroids. Newton was the first to prove that the net force is 0 (i.e., the
potential is constant) within these shells.  See \cite{cha87} for historical background. The infinitely thin homeoidal shell is a differential
element of a spheroid.  The homeoidal disk is the limiting case for which the minor axis approaches 0. The collapsed density of the thin homeoid is:

\begin{equation} \label{eq:HomSurfDens}
 \Sigma_{hom}(R;\alpha)  =   \begin{cases} {\sigma_0} /{\sqrt{1-R^2/\alpha^2}}  & \textrm{for } R<\alpha \\
                             0                                        & \textrm{for } R>\alpha
\end{cases}\end{equation}

\cite{lyn89} gives a formula for the external potential of the thin homeoidal shell.  Taking the limit $c=(1-e^2)^{1/2} \alpha\rightarrow0$ gives a
solution for the disk which is valid at all $R$ and $z$.

\cite{cud93} gives an expression for the potential of this disk which is simpler than previous solutions:
\begin{equation} \label{eq:CudPhi}
  \Phi_{hom}(R,z;\alpha) =
  -2\pi\alpha\sigma_0\G\arcsin\left[\frac{2\alpha}{\sqrt{z^2+(R+\alpha)^2} +  \sqrt{z^2+(R-\alpha)^2} }\right]
\end{equation}
Equation \ref{eq:CudPhi} can be further simplified. First make the trivial transformation:
\begin{equation} \label{eq:CudPhi1}
   \Phi_{hom}(R,z;\alpha) =
   -2\pi\alpha\sigma_0\G\arcsin\left[\frac{\sqrt{z^2+(R+\alpha)^2} -  \sqrt{z^2+(R-\alpha)^2}}{2R} \right]
\end{equation}
Now use the identity \ref{eq:IdentA1} to obtain
\begin{equation} \label{eq:HomPhi}
 \Phi_{hom}(R,z;\alpha) =
   -\pi\alpha\sigma_0\G~\left[\arcsin\left(\frac{ \alpha-\I z}{R}\right)+\arcsin\left(\frac{\alpha+\I z}{R}\right) \right]
\end{equation}

Equation \ref{eq:HomPhi} can be integrated over $\alpha$ whereas Equation \ref{eq:CudPhi1} yields an impossible integral. Equation \ref{eq:HomPhi}
must be real valued on physical grounds. This is easy to prove by noting that \arcsin(x) is an odd function of x and so the odd powers of $\I z$ in
the series expansion of equation \ref{eq:HomPhi} cancel, leaving a real valued result.

The gravitational field for the collapsed homeoid is obtained from the potential. Equation \ref{eq:HomPhi} yields particularly simple expressions for
the field vectors $ F_{R,hom}$ and $ F_{z,hom}$:
\begin{align}
 F_{R,hom}(R,z;\alpha)  &= -\frac{\pi\alpha\sigma_0\G}{R} \left[   \frac{\alpha-\I z}{\sqrt{R^2 - (\alpha- \I z)^2} }  +  \frac{\alpha+\I z}{\sqrt{R^2 + (\alpha -\I z)^2}
 }\right]\\
 F_{z,hom}(R,z;\alpha)  &= -{\pi\alpha\sigma_0\G} \left[   \frac{\I}{\sqrt{R^2-(\alpha-\I z)^2}}  -  \frac{\I}{\sqrt{R^2 - (\alpha+\I z)^2} }\right]
\end{align}

These force vectors can be expressed as entirely real functions using identities \ref{eq:IdentA8} and \ref{eq:IdentA9}:
\begin{align}
 F_{R,hom}(R,z;\alpha) &= -{\sqrt{2}\pi\alpha\sigma_0\G}~\frac{\alpha\sqrt{f_1 f_2 -f_3} - \abs{z}\sqrt{f_1 f_2 +f_3}}{ R f_1 f_2}\\
 F_{z,hom}(R,z;\alpha) &= -{\sqrt{2}\pi\alpha\sigma_0\G}~\frac{\sgn(z)\sqrt{f_1 f_2 +f_3}                            } {f_1 f_2 }
\end{align}
Where
\begin{align}\label{eq:Define-f1-f2-f3}
   f_1 &= \sqrt{z^2+(R +\alpha)^2}  \nonumber  \\
   f_2 &= \sqrt{z^2+(R -\alpha)^2}             \\
   f_3 &= \alpha^2 -R^2 -z^2        \nonumber \end{align}

\section{The n=1 Disk: The Maclaurin  disk } \label{sec:MaclaurinDisk}

Beginning in the early 18'th century Colin Maclaurin, along with James Ivory and many others, studied the properties of elliptical bodies.
\cite{cha87} includes a very good historical summary. See also \cite{bin08,ber00,sch56,mih68,kal71,kal72}.

The homogeneous oblate spheroid is the simplest case of a spinning body for which the gravitational attraction balances the centrifugal force. The
Maclaurin disk, also known as the Kalnajs disk \citep{kal72}, is a limiting case for which minor axis is 0. The Maclaurin disk is defined by the
surface density:
\begin{equation}\label{eq:MacSurfDens}
 \Sigma_{Mac}(R;\alpha)  =   \begin{cases}  \sigma_0 \sqrt{1-R^2/\alpha^2}  & \textrm{for } R<\alpha \\
                             0                                                          & \textrm{for } R>\alpha
\end{cases}\end{equation}

There are a few solutions for the potential of the Maclaurin disk in the literature. \cite{mih68} gives an expression for the potential of an oblate
homogeneous spheroid based on the derivation in \cite{sch56}. The potential of the Maclaurin disk can be found by letting the eccentricity  $ e
\rightarrow 1$ while holding the mass constant.  \cite{hun63} gives the solution for the Maclaurin disk as a series of Legendre polynomials in
elliptical coordinates. \cite{neu95,mei01,gon06} give a closed form solution for the potential of the Maclaurin disk in elliptic coordinates.

The starting point here is the potential-density pair for the n=0 disk, the flattened homeoid for which
$\Sigma_{hom}(R;\alpha)=\sigma_0/\sqrt{1-R^2/\alpha^2}$. The surface mass density of the Maclaurin disk is found from the transformation:
\begin{equation}
      \Sigma_{Mac}(R;\alpha) = \frac{1}{\alpha}  \int^\alpha_0{ \Sigma_{hom}(R; \hat{\alpha} )   ~d\hat{\alpha}}
 =   \frac{1}{\alpha}  \int^\alpha_0{\frac{\sigma_0}{\sqrt{1-R^2/\hat{\alpha}^2} } ~d\hat{\alpha}}= \sigma_0\sqrt{1-R^2/\alpha^2}
\end{equation}

The corresponding potential is:
\begin{equation}
    \Phi_{Mac}(R,z;\alpha) = \frac{1}{\alpha} \int^{\alpha}_0{ \Phi_{hom}(R;\hat{\alpha}) ~d\hat{\alpha}}
                           = \frac{-\pi \sigma_0 \G }{\alpha} \int^{\alpha}_0 {   \hat{\alpha} \left[\arcsin\left(\frac{\hat{\alpha}- \I z }{R}\right)+\arcsin\left(\frac{\hat{\alpha}+ \I z }{R}\right) \right]   d\hat{\alpha}}
\end{equation}
 Where the expression for $\Phi_{hom}$ is given by equation \ref{eq:HomPhi}~ above.  Use integral 2.813 and 2.833 from  \cite{gra94} to obtain:

\begin{equation}\label{eq:MacPhi-Im}
     \begin{split}
       \Phi_{Mac}(R,z;\alpha)&= -\frac{\pi\sigma_0 \G}{4\alpha}\bigg[
                               ( 2\alpha^2  -R^2 +2z^2  )\left(\arcsin\left(\frac{\alpha+\I z}{R}\right)+\arcsin\left(\frac{\alpha-\I z}{R}\right)\right)\\
                               &+\alpha      \left( \sqrt{R^2-(\alpha-\I z)^2} +\sqrt{R^2-(\alpha+\I z)^2} \right)\\
                               &-3\ z      \left( \I\sqrt{R^2-(\alpha-\I z)^2} -\I\sqrt{R^2-(\alpha+\I z)^2} \right)\bigg]
     \end{split}
\end{equation}

%
%
Equation \ref{eq:MacPhi-Im} can be converted to an entirely real expression by using the identities \ref{eq:IdentA1}, \ref{eq:IdentA6} and
\ref{eq:IdentA7}:
\begin{equation} \label{eq:MacPhi-Re}
     \begin{split}
        \Phi_{Mac}(R,z;\alpha) =-\frac{\pi\sigma_0 \G}{4\alpha}\bigg[&
                                2( 2\alpha^2 -R^2 +2z^2)\arcsin\left( \frac{f_1-f_2}{2 R} \right)\\
                                &+\sqrt{2}\alpha     \sqrt{f_1 f_2 -f_3}
                                -3\sqrt{2}\abs{z}  \sqrt{f_1 f_2 +f_3}  \bigg]
    \end{split}
\end{equation}
  where  $f_1, f_2, f_3$ are given by equation \ref{eq:Define-f1-f2-f3} above.

The gravitational field of the Maclaurin disk is found from the potential using $\Phi$ as given by equation \ref{eq:MacPhi-Im}. The resulting
expressions were expressed as entirely real functions by using the identities  \ref{eq:IdentA1}, \ref{eq:IdentA8}, and \ref{eq:IdentA9} :
\begin{equation}\label{eq:MacFR-Re}\begin{split}
F_{R,Mac}(R,z;\alpha)=& -\frac{\pi\sigma_0\G}{2 R \alpha} \Bigg[
   2 R^2\arcsin\left( \frac{f_1-f_2}{2 R} \right)\\
    & +\sqrt{2}\alpha  ( \alpha^2 -R^2 +z^2)   \frac{\sqrt{f_1 f_2 -f_3}}{f_1 f_2}\\
    & -\sqrt{2}\abs{z} ( \alpha^2 +R^2 +z^2)   \frac{\sqrt{f_1 f_2 +f_3}}{f_1 f_2}
  \Bigg]   \end{split}
\end{equation}
\begin{equation} \label{eq:MacFz-Re}\begin{split}
  F_{z,Mac}(R,z;\alpha)  = & -\frac{\pi\sigma_0 \G}{ \alpha} \Bigg[
     -2z\arcsin\left( \frac{f_1-f_2}{2 R} \right) \\
    & + 2 \sqrt{2} \alpha z \frac{  \sqrt{f_1 f_2 -f_3}}{f_1 f_2} \\
    & +\sqrt{2}~\sgn(z)~( \alpha^2 -R^2 -z^2) \frac{ \sqrt{f_1 f_2 +f_3}}{f_1 f_2}
 \Bigg]   \end{split}\end{equation}
       where  $f_1, f_2, f_3$ are given by equation \ref{eq:Define-f1-f2-f3} above.

The potential on the $z$ axis and on the $z=0$ plane can be found by by taking limits of equation \ref{eq:MacPhi-Re}:
\begin{equation}\label{eq:MacPhizaxis}
     \Phi_{Mac}( 0,z;\alpha) =  - \frac{\pi\sigma_0\G}{\alpha}\left[ \left( \alpha^2+ z^2  \right)\arcsin\left( \frac{\alpha}{\sqrt{\alpha^2+z^2}}   \right)
     -\alpha \abs{z} ~\right]
\end{equation}

\begin{equation}\label{eq:MacPhiOnDisk}
    \Phi_{Mac}(R, 0;\alpha) =    \begin{cases}-\frac{\pi^2\sigma_0\G}{4\alpha} \left( 2\alpha^2-R^2\right) & \textrm{for} R \leq \alpha       \\
                                              -\frac{\pi  \sigma_0\G}{2\alpha} \left[ (2\alpha^2-R^2)\arcsin(\frac{\alpha}{R})  +\alpha\sqrt{R^2-\alpha^2}\right]    &\textrm{for} R \geq \alpha
  \end{cases}
\end{equation}

The radial force vector in the $z=0$ plane is found by taking the limit of equation \ref{eq:MacFR-Re} or by differentiating equation
\ref{eq:MacPhiOnDisk} with respect to $R$.

\begin{equation}\label{FRMac_in_plane}
   F_{R,Mac}(R, 0;\alpha) =
         \begin{cases}  - \frac{\pi^2 R\sigma_0\G} {2 \alpha}& \textrm{for} R\leq\alpha\\
                        - \frac{\pi\sigma_0\G} {\alpha}\left[  {R\arcsin({\alpha}/{R}} )-\alpha\sqrt{1-\alpha^2/R^2}\right] & \textrm{for} R\geq\alpha
         \end{cases}
\end{equation}

The axial force vector on the $z$ axis is found by taking the limit of equation \ref{eq:MacFR-Re} or differentiating equation \ref{eq:MacPhizaxis}
with respect to $z$.

\begin{equation}\label{FzMac_in_plane}
   F_{z,Mac}( 0,z ;\alpha) =     -\frac{2\pi\sigma_0\G} {\alpha}\left[z\arcsin\left( \frac{\alpha}{\sqrt{\alpha^2+z^2}}\right) -\alpha~\sgn(z)\right]
\end{equation}

\section{The n=2  Disk } \label{sec:NewDisk}
 \cite{gon06} give a closed form solution for the potential of the n=2 disk in elliptic coordinates.

The disk surface density of the n=2 disk is
\begin{equation} \label{eq:D2SurfDens}
 \Sigma_{D2}(R;\alpha)  =   \begin{cases}  {\sigma_0} { (1-R^2/\alpha^2)^{3/2}}  & \textrm{for } R<\alpha \\
                                                               0                                         & \textrm{for } R>\alpha
\end{cases}\end{equation}

This mass distribution can be obtained from equation \ref{eq:MacSurfDens}, the disk surface density of the n=1 disk, with the transformation:
\begin{equation} \Sigma_{D2}(R;\alpha) =
    \frac{3}{\alpha^3}  \int^\alpha_0{ \hat{\alpha}^2  \Sigma_{Mac}(R; \hat{\alpha} )   d\hat{\alpha}}
\end{equation}

The corresponding potential is:
\begin{equation} \label{eq:D2PhiTransform}
    \Phi_{D2}(R,z;\alpha) = \frac{3}{\alpha^3}  \int^{\alpha}_0{ \hat{\alpha}^2 \Phi_{Mac}(R,z;\hat{\alpha}) ~d\hat{\alpha}}
\end{equation}
Where the expression for $\Phi_{Mac}$ is given by equation \ref{eq:MacPhi-Im} above.  Equation \ref{eq:D2PhiTransform} contains terms which after
change of variable have with the form  $ \int{x^n\arcsin(x) dx }$ and $\int{x^n\sqrt{1\pm x^2} dx}$ and can be solved using \cite{gra94} 2.262,
2.813, and 2.833.  After collecting terms, the result is reasonably compact:
\begin{small}
 \begin{equation}\label{eq:D2Phi-Im} \begin{split}
    \Phi_{D2}(R,z;\alpha) &= -\frac{\pi\sigma_0\G}{64\alpha^3}\Bigg[
                            3( 8{\alpha}^4 -8{\alpha}^2{R}^2 +16\alpha^2{z}^2 +3{R}^4 -24{R}^2{z}^2 +8{z}^4 ) \left(\arcsin\left(\frac{\alpha+\I z}{R}\right)+\arcsin\left(\frac{\alpha-\I z}{R}\right)\right) \\
                          &+\alpha ( 18\alpha^2 -9{R}^2 +26{z}^2 )    \left(  {\sqrt{R^2-(\alpha+\I z)^2}}  + {\sqrt{R^2-(\alpha-\I z)^2}}\right) \\
                          &-z(58\alpha^2 -55{R}^2 +50{z}^2 )     \left( {\I}{\sqrt{R^2-(\alpha+\I z)^2}}  - {\I}{\sqrt{R^2-(\alpha-\I z)^2}}\right)
    \bigg]
\end{split} \end{equation} \end{small}

Equation \ref{eq:D2Phi-Im} can be converted to an entirely real expression by using the identities \ref{eq:IdentA1}, \ref{eq:IdentA6} and
\ref{eq:IdentA7}:
\begin{equation}\label{eq:D2Phi-Re}\begin{split}
  \Phi_{D2}(R,z;\alpha) =&-\frac{\pi\sigma_0\G}{64\alpha^3}\Bigg[
       6( 8{\alpha}^4 -8{\alpha}^2{R}^2 +16\alpha^2{z}^2 +3{R}^4 -24{R}^2{z}^2 +8{z}^4 )\arcsin\left( \frac{f_1-f_2}{2 R}   \right)\\
      &+\sqrt{2}\alpha  ( 18\alpha^2 -9{R}^2 +26{z}^2 )   \sqrt{f_1 f_2 -f_3}\\
      &-\sqrt{2}\abs{z} (58\alpha^2 -55{R}^2 +50{z}^2 )            \sqrt{f_1 f_2 +f_3}
  \Bigg]
\end{split} \end{equation}
       where  $f_1, f_2, f_3$ are given by equation \ref{eq:Define-f1-f2-f3} above.

The gravitational field of the n=2 disk is the gradient of the potential using $\Phi$ as given by equation \ref{eq:D2Phi-Im}.  The resulting
expressions were expressed as entirely real functions by using the identities by using identities \ref{eq:IdentA1}, \ref{eq:IdentA8}, and
\ref{eq:IdentA9} :

\begin{small}
\begin{equation}\label{eq:D2FR-Re} \begin{split}
   F_{R,D2}(R,z;\alpha)  &= - \frac{3\pi\sigma_0\G}{16 R \alpha^3}  \Bigg[
         2R^2 (4\alpha^2R^2 -3R^4  +12 R^2z^2 )\arcsin\left( \frac{f_1-f_2}{2 R}   \right) \\
      &+ \sqrt{2}\alpha(  2\alpha^4 -5\alpha^2R^2 +4\alpha^2z^2 +3R^4 -25R^2z^2 +2z^4 )  \frac{ \sqrt{f_1 f_2 -f_3}}{f_1 f_2} \\
      &+ \sqrt{2}\abs{z}( 2\alpha^4 +9\alpha^2R^2 +4\alpha^2z^2 -13R^4 -11R^2z^2 +2z^4 )    \frac{ \sqrt{f_1 f_2 +f_3}}{f_1 f_2}
                \Bigg]
\end{split}\end{equation} \end{small}
\begin{small}
\begin{equation}\label{eq:D2Fz-Re} \begin{split}
   F_{z,D2}(R,z;\alpha)   & = - \frac{\pi\sigma_0\G}{4 \alpha^3}  \Bigg[
        -6z( 2\alpha^2  -3R^2 +2z^2 )\arcsin\left( \frac{f_1-f_2}{2 R} \right) \\
      & + \sqrt{2} \alpha z ( 13\alpha^2 -13R^2 +17z^2  )          \frac{ \sqrt{f_1 f_2 -f_3}}{f_1 f_2} \\
      & + \sqrt{2} ~\sgn(z) (( 4\alpha^4 -8\alpha^2R^2 -3\alpha^2z^2 +4R^4 -7R^2z^2 -11z^4 )   \frac{ \sqrt{f_1 f_2 +f_3} } {f_1 f_2}
                \Bigg]
 \end{split}\end{equation} \end{small}
       where  $f_1, f_2, f_3$ are given by equation \ref{eq:Define-f1-f2-f3} above.

The potential on the $z$ axis and on the $z=0$ plane can be found by by taking the limit of equation \ref{eq:D2Phi-Re}:
\begin{equation}\label{eq:D2Phizaxis}
     \Phi_{D2}( 0,z;\alpha) =  - \frac{\pi\sigma_0\G}{4\alpha^3}\left[ 3\left( \alpha^4+2\alpha^2 z^2 + z^4  \right)\arcsin\left(\frac{\alpha}{\sqrt{\alpha^2+z^2}}\right)
     -\alpha \abs{z}(5\alpha^2+3z^2) ~\right]
\end{equation}

\begin{equation} \label{eq:D2PhiOnDisk}
    \Phi_{D2}(R,0;\alpha) =    \begin{cases}-\frac{3\pi^2\sigma_0\G}{64\alpha^3} \left( 8\alpha^4 -8\alpha^2R^2+3R^4\right) & \textrm{for} R \leq \alpha       \\
                                              -\frac{3\pi\sigma_0\G}{32\alpha} \left[ ( 8\alpha^4 -8\alpha^2R^2+3R^4)\arcsin(\frac{\alpha}{R})
+3\alpha(2\alpha^2-R^2)\sqrt{R^2-\alpha^2}\right]    &\textrm{for} R \geq \alpha
  \end{cases}
\end{equation}

The radial force vector in the $z=0$ plane is found by taking the limit of equation \ref{eq:D2FR-Re} or differentiating equation \ref{eq:D2PhiOnDisk}
with respect to $R$.

\begin{equation}\label{FRD2_in_plane}
   F_{R,D2}(R, 0;\alpha) =
         \begin{cases}  -\frac{3\pi^2 R\sigma_0\G} {16\alpha^3}(4\alpha^2-3R^2)& \textrm{for} R\leq\alpha\\
                        -\frac{3\pi\sigma_0\G} {8\alpha^3}\left[ R(4\alpha^2-3R^2)\arcsin({\alpha}/{R} )
                                           -\alpha(2\alpha^2-3R^2)\sqrt{1-\alpha^2/R^2}\right] & \textrm{for} R\geq\alpha
         \end{cases}
\end{equation}

The axial force vector on the $z$ axis is found by taking the limit of equation \ref{eq:D2Fz-Re} or differentiating equation \ref{eq:D2Phizaxis} with
respect to $z$.

\begin{equation}  \begin{split}\label{eq:FzD2_OnZaxis}
   F_{z,D2}( 0,z ;\alpha) = - \frac{\pi\sigma_0\G}{2 \alpha^3( \alpha^{2}+{z}^{2}) } \bigg[
  & -6(\alpha^2+z^2)^2 z \arcsin\left(\frac{\alpha}{\sqrt{\alpha^2+z^2}}\right)\\
  & +\alpha z^2(13\alpha^2+17z^2)
  +\alpha\,\sgn(z)(4\alpha^4-3\alpha^2z^2-11z^4)
  \bigg]  \end{split}\end{equation}

\subsection{Comparison of the Maclaurin disk and the n=2 disk}

\begin{figure*}[t]
  \plottwo{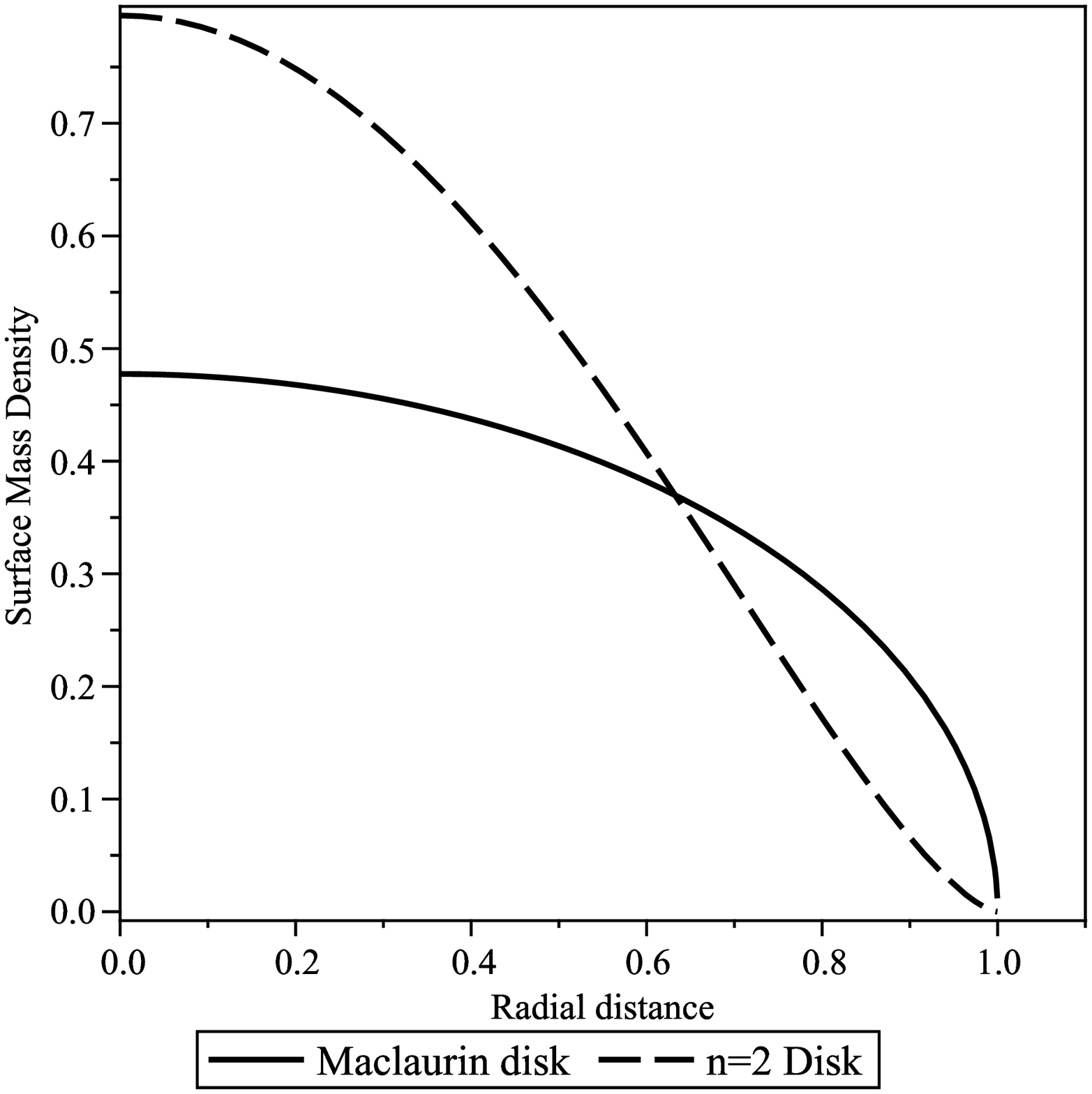}{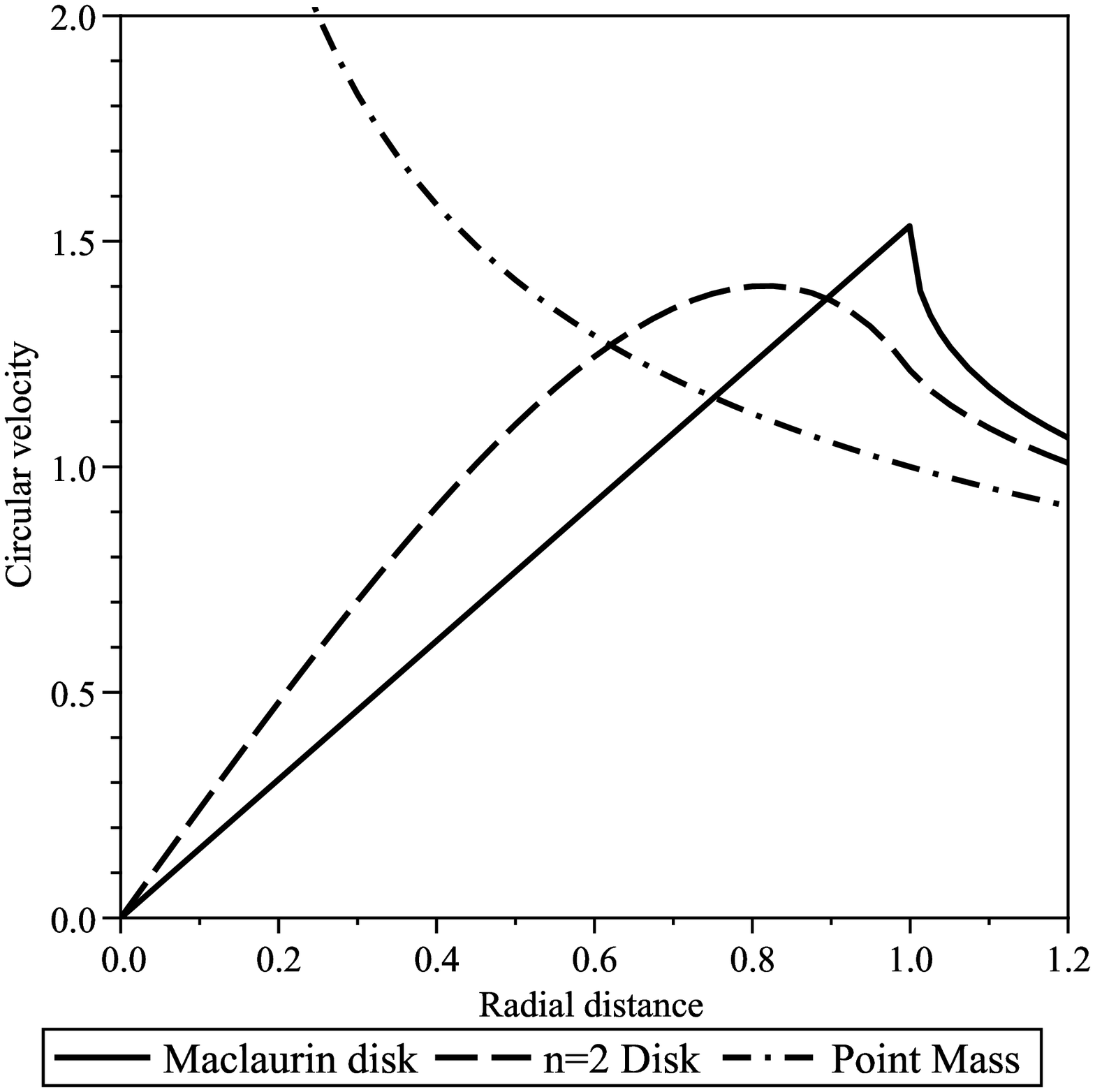}
  \caption{Compare the surface mass density and circular velocity, $V_c=\sqrt{(-V F_R(R,z)}$, for the Maclaurin disk and the n=2 disk.  Reduced units are used so that
           the masses of both disks and the point mass are 1.0 }
  \label{fig:Sigma+Velocity}
\end{figure*}

Table 1 compares important properties of the two disks.  Figure \ref{fig:Sigma+Velocity}a compares the surface mass density and potential.  Figure
\ref{fig:Sigma+Velocity}b compares rotational velocity in the disks.  The n=2 disk is more centrally concentrated than the Maclaurin disk. The
rotational velocity increases more quickly in the inner disk and begins to fall before reaching the edge of the disk.  As is apparent from figure
\ref{fig:Sigma+Velocity}b, the derivative of the circular velocity of the Maclaurin disk is discontinuous at the edge of the disk whereas the n=2
disk is better behaved.

\renewcommand{\arraystretch}{1.5}
\begin{tabular} { l  r l l}  \\
\multicolumn{3}{c}{  \bfseries Table 1 comparison of the Maclaurin and the n=2 disks}\\
\bfseries ~~Property&&\bfseries ~~Maclaurin disk~~~&\bfseries ~~n=2 Disk\\
  \hline
     Surface density       &$ \Sigma(R)         =$&$ \sigma_0\sqrt{1 - {R^2}/{\alpha^2}}    $&$= \sigma_0 (1 - {R^2}/{\alpha^2})^{3/2}                         $\\
     Total mass            &$ M                 =$&$ \frac23 \pi\alpha^2\sigma_0            $&$= \frac25\pi \alpha^2\sigma_0                                  $\\
      Circular velocity    &$ V_c^2(R,0)        =$&$ \dfrac{\pi^2 R^2\sigma_0\G} {2\alpha}  $&$= \dfrac{3\pi^2 R^2\sigma_0\G (4\alpha^2 -3 R^2) } {16\alpha^3} $\\
                                               &= &$ \dfrac{3\pi R^2 M\G}{4\alpha^3}        $&$= \dfrac{15\pi R^2 M \G(4\alpha^2 -3 R^2)}{32\alpha^3}          $\\
     Disk edge velocity    &$ V_c^2(\alpha,0)   =$&$ \dfrac{\pi^2 \alpha \sigma_0\G} {2}    $&$= \dfrac{3\pi^2 \alpha \sigma_0\G} {16}                         $\\
                                               &= &$ \dfrac{3 \pi M\G}{4\alpha}             $&$= \dfrac{15\pi M\G}{32\alpha}                              $\\
  \hline
\end{tabular}

\section{Example: The force field of a simple galaxy model}\label{sec:apps}

\begin{figure*}[t]
\epsscale{.50}
  \plotone{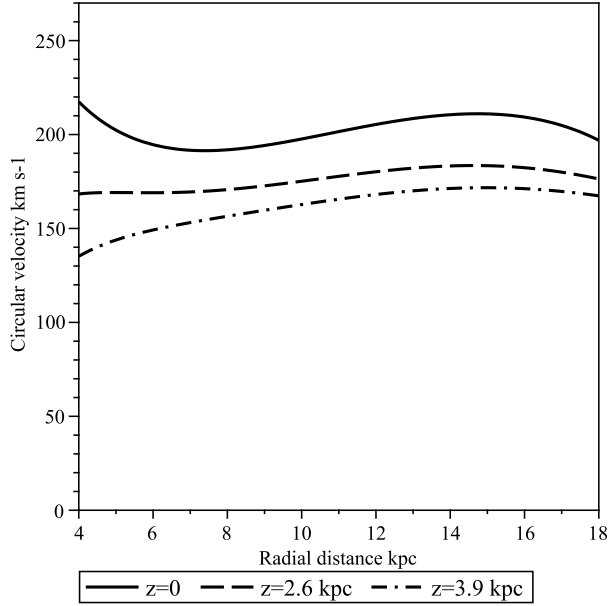}
  \caption{ Circular velocity, $V_c=\sqrt{(-V F_R(R,z)}$, of a simple galaxy model at z=0, z=2.6~\kpc, and z=3.9~\kpc\, above the disk. The model consists of an n=2 disk and a
           point mass to represent the core + bulge. The masses of the two components are $M_{disk}=3.5\dex{10} \Msun$ and
           $M_{Point Mass} = 4\dex{10} \Msun$.  The radius of the disk is $\alpha=19~\kpc$.  }
  \label{fig:NGC891}
\end{figure*}

Three-dimensional problems such as that of the structure and kinematics of the extra-planar gas will benefit from the use of the new
density-potential pairs.  A simple galaxy model was constructed for illustration.  The model consists of a an n=2 disk and a core/bulge region
modeled as a point mass.  This model is defined by three parameters: the mass of the disk, the mass of the core/bulge region, and the diameter of the
disk. As shown in figure \ref{fig:NGC891}, the circular velocity, calculated as $V_c=\sqrt{(-V F_R(R,z)}$, is nearly constant over much of the disk.
Also, the derivative of the circular velocity with $z$ is nearly linear over a wide range of both $R$ and $z$.

Figure \ref{fig:NGC891} agrees surprisingly well with figure 5 of \cite{fra05::1} which shows that the measured velocity of HI for NGC891 decreases
linearly with the height above the disk.  See also  \cite{ran97,swa97,kam07,oos07,fra06::3,bar06}. Further work is planned on this topic.

\section{Summary and conclusion}

We have presented new solutions for a family of finite disks.  Closed form expressions in cylindrical coordinates using elementary functions are
given for the potential and gravitational force for the disks with surface density $\Sigma_n=\sigma_0 (1-R^2/\alpha^2)^{n-1/2} \textrm{ with } n=0,
1, 2$.  Expressions are also given for the limiting cases of $R=0$ and $z=0$.

These solutions fill a need and should make it easier to model 3-D gravitational phenomenon involving disk galaxies.  This is particularly important
due to the recent availability of detailed kinematic data above the plane of the disk.

\acknowledgements

I am grateful to the anonymous referee for useful suggestions and comments which improved the presentation.

\newpage
\appendix

\section{Some simple identities}
\label{app-identities}

A number of identities are gathered here.  In all cases $ x, y \in \Re$  The ranges of validity must avoid the discontinuity of the principal value
of the square root function on the negative real axis.

Equation \ref{eq:IdentA1} can be proved by taking the sin of both sides; reducing the terms using the identities  $\sin(a+b) =
\sin(a)\cos(b)+\cos(a)\sin(b)$ and $\sin(2a) = 2\sin(a)\cos(b)$; and substituting $\cos = \sqrt{1-\sin^2}$. The other identities can be found by
substituting into the relation
 $$\sqrt{x+\I y}=\frac{  \sqrt{ \sqrt{x^2+y^2} + x} +  \sgn(y) \I \sqrt{ \sqrt{x^2+y^2} - x}} {\sqrt{2}}$$
and into the expression found by taking the reciprocal of both sides.
\begin{small}
  \begin{align}
   \label{eq:IdentA1}\arcsin(x-\I y) +\arcsin\left(x+\I y\right)           &~=~  2\arcsin\left[ \onehalf\sqrt{(x+1)^2+y^2} - \onehalf\sqrt{(x-1)^2+y^2}\right]\\
   \label{eq:IdentA2}  \sqrt{x-\I y}+\sqrt{x+\I y}                          &~=~  \sqrt{2}\sqrt{ \sqrt{x^2+ y^2} +x}\\
   \label{eq:IdentA3}  \I\sqrt{x-\I y} -  \I\sqrt{x+\I y}                   &~=~  \sgn(y) \sqrt{2} \sqrt{ \sqrt{x^2+ y^2} -x}\\
   \label{eq:IdentA4}  \frac{1}{\sqrt{x+\I y}}+ \frac{1}{\sqrt{x-\I y}}     &~=~  \sqrt{2}\frac{ \sqrt{\sqrt{x^2+ y^2}+x}}{\sqrt{x^2+ y^2}  }\\
   \label{eq:IdentA5}  \frac{\I}{\sqrt{x+\I y}} -\frac{\I}{\sqrt{x-\I y}}   &~=~
                \sgn(y)\sqrt{2} \frac{\sqrt{ \sqrt{x^2+ y^2} -x} } {\sqrt{x^2+ y^2}}\\
   \label{eq:IdentA6}  {\sqrt{1- \left( x+\I y \right) ^2}}+ {\sqrt{1-\left( x-\I y \right) ^2}}&~=~
                 \sqrt{2} {\sqrt{  \sqrt{1 -2x^2 +2y^2+x^4+2x^2y^2 +y^4} +1 -x^2 +y^2}} \\
   \label{eq:IdentA7}   \I  {\sqrt{1- \left( x+\I y \right) ^2}}-  {\I }{\sqrt{1-\left( x-\I y \right) ^2}} &~=
                  \sgn(xy)\sqrt{2} {\sqrt{ \sqrt{1 -2x^2 +2y^2+x^4+2x^2y^2 +y^4} -1 +x^2 -y^2}} \\
   \label{eq:IdentA8}  \frac{1}{\sqrt{1- \left( x-\I y \right) ^2}}+\frac {1}{\sqrt{1-\left( x+\I y \right) ^2}}&~=~
                \sqrt{2}\frac{\sqrt{  \sqrt{1 -2x^2 +2y^2+x^4+2x^2y^2 +y^4} +1 -x^2 +y^2}}
                {\sqrt{1 -2x^2 +2y^2+x^4+2x^2y^2 +y^4}}\\
   \label{eq:IdentA9}  \frac{\I }{\sqrt{1- \left( x-\I y \right) ^2}}- \frac {\I }{\sqrt{1-\left( x+\I y \right) ^2}} &~=
                \sgn(xy)\sqrt{2}\frac{\sqrt{  \sqrt{1 -2x^2 +2y^2+x^4+2x^2y^2 +y^4} -1 +x^2 -y^2}}
                {\sqrt{1 -2x^2 +2y^2+x^4+2x^2y^2 +y^4}}
 \end{align}
\end{small}

\newpage


\end{document}